\font\twelvebf=cmbx12
\font\ninerm=cmr9
\nopagenumbers
\magnification =\magstep 1
\overfullrule=0pt
\baselineskip=18pt
\line{\hfil }
\line{\hfil May 1997}
\vskip .8in
\centerline{\twelvebf  Matrix String Theory As A Generalized Quantum 
Theory}
\vskip .5in
\centerline{\ninerm D.MINIC}
\centerline{Physics Department}
\centerline{Pennsylvania State University}
\centerline{University Park, PA 16802}
\centerline{and}
\centerline{Enrico Fermi Institute}
\centerline{University of Chicago}
\centerline{Chicago, IL 60637}
\centerline {dminic@yukawa.uchicago.edu}

\vskip 1in
\baselineskip=16pt
\centerline{\bf Abstract}
\vskip .1in

Matrix String Theory of Banks, Fischler, Shenker
and Susskind can be understood 
as a generalized quantum theory (provisionally named
"quansical" theory) which differs from
 Adler's generalized trace quantum dynamics. The effective
planar Matrix String Theory Hamiltonian 
is constructed in a particular fermionic realization of
Matrix String Theory treated as an example of "quansical" theory.

\vfill\eject

\footline={\hss\tenrm\folio\hss}

\magnification =\magstep 1
\overfullrule=0pt
\baselineskip=22pt
\pageno=2
\noindent{\bf 1. Introduction}
\vskip .2in

In this article I want to show how a generalized
quantum-theoretical structure (in which $\hbar$ is kept finite,
so that the classical limit is not defined by letting 
$\hbar \rightarrow 0$) naturally appears
 in connection with the problem of non-perturbative
formulation of string theory.  
Matrix String Theory (the planar SUSY Yang-Mills quantum mechanics
of Polchinski's D0-branes [1] [2] [3], the suggested partonic
constituents of the fundamental strings) presents one such formulation. 

How can one go about formulating 
the classical limit of a quantum theory while keeping $\hbar$ finite?
One possible approach was put forward by Adler [4].
The idea is to avoid "quantization" altogether and directly formulate 
Hamiltonian dynamics on a non-commuting phase space for general 
non-commuting degrees of freedom. Adler assumes that operator 
multiplication
of non-commuting operator variables is associative
and that there exists a graded trace ${\bf Tr}$ obeying
the fundamental property of cyclic permutation of non-commuting
operator variables according to 
${\bf Tr} O_{1} O_{2} = \pm {\bf Tr} O_{2} O_{1}$, where
$\pm$ corresponds to the situation when both $O_{1}$ and $O_{2}$ are
bosonic/fermionic. Then it can be shown [4] 
that for a general trace functional
${\bf A}$, defined as the graded trace ${\bf Tr}$ 
of a bosonic polynomial function
of operator variables $q_{i}$, one can uniquely define 
${{\delta {\bf A}} \over {\delta q_{i}}}$, from 
${\delta {\bf A}} = {{\delta {\bf A}} \over {\delta q_{i}}} {\delta 
q_{i}}$.
Then Adler shows how to define the generalized Poisson bracket, 
formulate generalized Hamilton equations of motion, etc. [4].
The whole structure nicely applies to Matrix String Theory.
 
Now, given the fact that Matrix String Theory is the $N=\infty$ limit of 
the supersymmetric quantum mechanics describing $D0$-branes, it
is important to incorporate the basic features of the planar limit,
such as factorization, 
in order to be able to discuss any dynamical issues.
(Factorization by definition means that 
${\langle} F_{1} F_{2}{\rangle} = 
{\langle}F_{1}{\rangle}{\langle} F_{2}{\rangle}$ [5], given
two observables $F_{1}$ and $F_{2}$.)
Therefore, I propose to study a generalized quantum theory which naturally
contains certain features of classical dynamics, implied by
factorization. Such a theory can be induced 
starting from the Feynman-Schwinger differential variational principle 
[6][7]
$$
{\langle} F \delta_{X} S {\rangle} = i  \hbar  {\langle} \delta_{X} F 
{\rangle}.
\eqno(1)
$$
Here $S$ is the classical action, given in terms of some variables
$X$, and $F$ is some appropriate observable.
As is well-known, one can deduce 
the canonical Heisenberg
commutation relations (by choosing $F=X$), 
the quantum dynamical equations of motion etc., starting from (1) [6][7].
The correspondence principle is by definition manifested in the
classical nature of $S$. Bohr's complementarity principle  
is formally contained in the Heisenberg uncertainty relations,
that follow from the canonical commutation relation.

Suppose that one changes the meaning of $\delta_{X}$ in (1) in 
order to generate a generalized version of quantum theory
in which the classical property 
of factorization holds when the dynamical 
variables that appear in the variational
derivative happen to be non-commuting.
That would imply a summation over the planar trajectories only
in the path-integral, the integral version of (1). 
One would be then dealing with a quantum theory with certain classical 
features.
(One might call such a generalization of quantum theory, planar
quantum theory or "quansical" theory.) 

In what follows I wish to show that Matrix String Theory of Banks, 
Fischler,
Shenker and Susskind [1] can be understood as an example of
 "quansical" theory (see [8]
for an application of the same set of ideas to the planar Yang-Mills 
theory).
The Matrix String Theory Hamiltonian written
in terms of the nine non-commutative coordinates $X_{i}$, and their 
sixteen fermionic superpartners $\Phi$ reads
($R \rightarrow \infty$ defines the decompactified limit
of M theory, as in [1])
$$
H = R tr({1 \over 2} {P_{i}}^{2} + {1 \over 4} [X_{i},X_{j}]^{2} +
\Phi^{T} \gamma_{i} [\Phi, X_{i}]),
\eqno(2) 
$$
where $P_{i}$ is canonically conjugate to $X_{i}$.
In the following I also wish to present a particular heuristic 
realization 
of Matrix String Theory treated as
an example of "quansical" theory which should be useful for further
analyses of many important dynamical questions in Matrix String Theory.
The emphasis is placed on the planar nature of the theory. Possible 
subtleties
due to supersymmetry are not taken into consideration. 

\vskip .2in
\noindent{\bf 2. Planar Quantum Theory, Alias "Quansical" Theory}
\vskip .2in

First let me state the definition of planar quantum theory or
"quansical" theory. The Feynman-Schwinger differential variational 
principle 
(1) can be used to postulate the 
following Euclidean version of quantum dynamics
that is consistent with factorization [9]. (I concentrate on
the bosonic variables for the time being):
$$
({{\delta S}  \over {\delta X_{\mu}}}- 2 \Pi_{\mu}) |0{\rangle} = 0,
\eqno(3)
$$
where
$$
[X_{\mu},{\Pi}_{\nu}] =  \delta_{\mu \nu} \hbar |0{\rangle}{\langle}0|.
\eqno(4)
$$
Equations (3) and (4) simply represent 
the original equation (1) written in such
a way so that the property of factorization is valid, in other words, so
that the only state that survives the planar limit is the ground state.
Equations (3) and (4) define a Euclidean planar quantum theory, or 
Euclidean "quansical" theory.  

Likewise, one could consider a Hamiltonian version of the
same planar quantum theory (see again [9]). 
The "quansical" commutation relations are given by
$$
[X_{i},P_{j}] = i \delta_{ij} \hbar |0{\rangle}{\langle}0|.
\eqno(5)
$$
(In other words, in the expansion of unity that appears in the
usual canonical commutation relations $[X_{i},P_{j}] = i \delta_{ij} 
\hbar$
$$
1 = |0{\rangle}{\langle}0| + \sum_{n=1} |n{\rangle}{\langle}n|,
\eqno(6)
$$
only the first term, which is a projection operator, is kept.
Expression (5) can be taken as the starting point for a 
discussion of a generalized version of the Heisenberg uncertainty
principle, and therefore, a generalized version of Bohr's complementarity 
principle.)
The dynamical equations of motion are given by the familiar expressions
$$
i [H_{r},X_{i}]= \hbar \dot X_{i},
\eqno(7)
$$
and
$$
i [H_{r},P_{i}]= \hbar \dot P_{i}.
\eqno(8)
$$
It is important to note  
that $H_{r}$ represents the reduced Hamiltonian (reduced onto the
ground state of the theory). 
Equations (5), (7) and (8) define a Hamiltonian planar quantum theory, or
Hamiltonian "quansical" theory.
I wish to adopt the Hamiltonian version of "quansical" theory in the 
following 
discussion of Matrix String Theory.

\vskip .2in
\noindent{\bf 3. Fermionic Realization}
\vskip .2in

It seems apparent that it is absolutely crucial to 
come up with a suitable realization of the 
ground state in order to be able to use the above definiton of 
planar quantum theory. Consider then the following concrete realization 
of such  
generalized quantum Hamiltonian dynamics 
based on a very particular representation of the projection operator in 
the
"quansical" commutation relations (5):
$$
|0{\rangle}{\langle}0| = \psi^{\dagger} \psi,
\eqno(9)
$$
where $\psi^{2}={\psi^{\dagger}}^{2} = 0$, 
$\psi \psi^{\dagger} + \psi^{\dagger} \psi =1$, i.e. $\psi$ and
$\psi^{\dagger}$ are fermionic operators.
This representation is suggestive of a fermionic ground state.

Using the commutation relation (5) one deduces that, at least formally, 
$$
P_{j} = i \hbar X_{i}^{-1} (\delta_{ij} C +  
X_{i}^{-1} \delta_{ij} C X_{i} 
+ (X_{i}^{-1})^{2} \delta_{ij} C {X_{i}}^{2} + ...),
\eqno(10)
$$
where $C^{2}=C$ is the projection operator 
$|0{\rangle}{\langle}0| = \psi^{\dagger} \psi$.
Likewise,  
the following formal expression for $H_{r}$, in terms of $X_{i}$ and 
$P_{i}$,
stems from (7) ($P_{i} = \dot X_{i}$) 
$$
H_{r} = i \hbar X_{i}^{-1} ( P_{i} +  X_{i}^{-1} P_{i} X_{i} +
(X_{i}^{-1})^{2}  P_{i} {X_{i}}^{2} + ...).
\eqno(11)
$$
Therefore, given the particular realizations of the projection
operator $|0{\rangle}{\langle}0|$ and $X_{i}$ the relevant representations
of $P_{i}$ and $H_{r}$ follow from (10) and (11). The reduced (or 
effective)
planar Hamiltonian $H_{r}$ completely defines the dynamics of the planar
limit of a matrix theory under cosideration
(in the present context, Matrix String Theory).

So, apart from choosing a suitable realization for the ground state
(such as (9)) one has to pick a particular representation for $X_{i}$, 
that is
compatible with the already chosen realization for the ground state,
and then deduce the reduced planar Hamiltonian.

Suppose the following dictionary is used to 
postulate a particular realization of $X_{i}$ (and its supersymmetric 
partners $\Phi$): 
$$
X_{i} \rightarrow x_{i}(\alpha,\beta) ,
\Phi \rightarrow \phi(\alpha,\beta),
   \eqno(12)
$$ 
where $\alpha$ and $\beta$ are
two real parameters, and $x_{i}$'s and $\phi$'s
 are functions of $\alpha$, $\beta$.
(In other words, the matrix indices, which play the role of
 internal parameters
and which in the planar limit run from zero to
infinity, are replaced by two continuous, external parameters 
$\alpha$ and $\beta$.) 
Likewise, let the commutator bracket be replaced according to the
following prescription:
$$
[X_{i}, X_{j}] \rightarrow 
\{x_{i}(\alpha,\beta), x_{j}(\alpha,\beta)\},
[\Phi,X_{i}] \rightarrow \{\phi(\alpha,\beta),x_{i}(\alpha,\beta)\},  
\eqno(13)
$$
where $\{,\}$ denotes the ordinary Poisson bracket with respect to 
 $\alpha$ and $\beta$, in other words
$$
\{x_{i}(\alpha,\beta), x_{j}(\alpha,\beta)\} =
(\partial_{\alpha}x_{i}(\alpha,\beta)  
\partial_{\beta}x_{j}(\alpha,\beta) -
\partial_{\beta}x_{i}(\alpha,\beta)
\partial_{\alpha}x_{j}(\alpha,\beta)) .           \eqno(14)
$$
Thus all expressions containing 
the non-commutative coordinates
and the commutator are to be replaced with the "identically"
looking ones, after the translation defined
by (12) and (13) has been applied. Also, the operation of
tracing should be replaced by the operation of integration over the extra 
continuous parameters $\alpha$ and $\beta$
$$
Tr  \rightarrow   \int d\alpha d\beta.   \eqno(15)
$$

(This is essentially the dictionary of [10], where the equivalence
between the $SU(\infty)$ Lie algebra and the algebra of
area preserving diffeomorphisms of a two-dimensional sphere $S^{2}$,
parametrized by $\alpha$ and $\beta$, was presented.)

The $SU({\infty})$ structure constants translate according to 
$$
X_{i}^{c}t^{c}  \rightarrow 
\sum_{lm} x_{i}^{lm} 
Y_{lm}(\alpha,\beta) ,             \eqno(16)
$$
where $t^{c}$ are the generators of $SU(\infty)$ and 
$Y_{lm}(\alpha,\beta)$
are the $S^{2}$ spherical harmonics. The $SU(\infty)$ structure constants
are then identified with the structure constants of the area preserving 
diffeomorphisms of a two-sphere, defined in terms of the spherical 
harmonics basis [10]
$$
\{Y_{lm},Y_{l'm'}\} = f_{lm,l'm'}^{l''m''}Y_{l''m''}.     \eqno(17)
$$

The expression defining the $SU(\infty)$ gauge transformations (which
generalizes the usual coordinate transformations of commuting 
coordinates) 
reads, for example, as follows
$$
\delta x_{i}(\alpha, \beta) = 
\{x_{i},\Omega\}.                \eqno(18)
$$

The Matrix String Theory Hamiltonian, on the other hand, becomes
$$
H = R\int d\alpha d\beta ({1 \over 2}p_{i}^{2} + {1 \over 
4}\{x_{i},x_{j}\}^{2}
+  \phi^{T} \gamma_{i} \{\phi,x_{i}\}),
\eqno(19)
$$
where $p_{i} \rightarrow -i{\delta \over {\delta x_{i}}}$.
  
The above realization of $X_{i}$ is multiplicative, that is
$$
X_{i} f(x) = x_{i} f(x),
\eqno(20)
$$
where $f(x)$ is an appropriate functional of $x_{i}$. (Remember that
$x_{i}$'s are functions of $\alpha$ and $\beta$.)
Given that fact, it formally follows from (10) that
$$
P_{i} f(x) = \int Dz_{j}(\alpha, \beta) 
{{{\delta}_{ij}(z) \rho(z) f(z)} \over {x_{j}-z_{j}}}.
\eqno(21)
$$
Here, by definition, 
$\delta_{ij}(z) \rightarrow \delta_{ij}$ for $z \rightarrow x$.
and the scalar product of two wave functionals is defined as
$$
{\langle} f g {\rangle} = \int Dx_{i} \rho(x) f^{*}(x) g(x),
\eqno(22)
$$
where the weight functional $\rho$ 
satisfies the following constraint (as in [11] [12], even though
the present $\rho$ is not related to any density of eigenvalues)
$$
\int Dx_{i} \rho(x) = 1.
\eqno(23)
$$
Note that (23) implies that the 
ground wave functional is set to one, so that the
 ground state is completely described in terms of 
the weight functional $\rho$.
In other words, the physical nature of the
weight functional $\rho$ is that it could be taken as another
realization of the ground state. 
(By taking $\rho$ to be a constant equal to the inverse of $\int Dx_{i}$ 
one
can recover the canonical commutation relations and
the canonical scalar product of two wave functionals. From this
point of view the functional $\rho$ is a "deformation" 
functional responsible for the described generalization of ordinary
quantum theory into "quansical" theory.)
The action of the projector $|0{\rangle}{\langle}0|$ is given by
the following expression (in view of (22))
$$
\psi^{\dagger} \psi |f{\rangle} = \int Dx_{i} \rho (x) f(x).
\eqno(24)
$$
This particular relation can be taken to 
define a fermionic ground state of the theory.
(The same statement is familiar from
the original one-matrix model study [12], even though it is
clear that the present realization does not speak of any
eigenvalues.) More precisely, there exists
a Fermi surface, the Fermi energy playing the role of a Lagrange 
multiplier due to the constraint (23).

That can be seen from the
"quansical" commutation relations (5), given the fermionic
realization of the projector (9).
First note that
(5) implies 
${\langle}0|[X_{i},P_{j}]|0{\rangle}
 = i \delta_{ij} $, which in view of (12)
implies $P_{i} \rightarrow p_{i}(\alpha,\beta)$.
Then note that due to the fact that $\psi^{\dagger} \psi$ is a fermion
number operator, each phase cell $\Delta x_{i} \Delta p_{i}$ contains
a single fermion. (The same fermion number operator serves as a generator 
of
the area preserving diffeomorphisms of $S^{2}$, which is compatible with
(12) and (13).) 
The ground state is then essentially 
 characterized by a certain region of the functional
phase space $Dx_{i} Dp_{i} D\phi^{T} D \phi$ 
which possesses the property
of incompressibility according to the Liouville theorem. 
In other words the 
following constraint is valid
$$
\int Dx_{i} Dp_{i}  D\phi^{T} D\phi \theta(e - H) =1,  
 \eqno(25)
$$
where $H$ denotes the Hamiltonian (19), $e$ stands for the characteristic 
Fermi
energy and $\theta$ is the usual step function. 
 
Equation (25)
tells us that the volume of the functional phase space fluid is to be 
normalized to one in such a way, as if there existed a single 
fermion placed at each phace space
cell, and consequently, taking into account 
the Pauli exclusion principle, as if there existed, in the limit
of a large number of cells,  
an incompressible fermionic fluid, with
the Fermi energy $e$. 
By recalling that each phace space cell has 
a natural volume of the order of the Planck constant and that 
the planar
limit corresponds to a situation where the number of cells
goes to infinity, the product of the Planck
constant and the number of cells can be adjusted to one (the reason being 
that the $1/N$ expansion formally 
corresponds to a "semiclassical" expansion, $1/N$ acting as an effective
"Planck constant"). Then follows the
relation (25), describing an incompressible drop
of functional phase space of unit volume. 
(Note that the appearance of fermions 
could be intuitively understood from the point of view of 't Hooft's
 double-line
representation for the planar graphs [5]. The fact that such
 lines do not
cross in the planar limit is achieved by attaching fermions to each line 
and
using the exclusion principle.) 

Therefore, equation (25) gives a rather natural, even though
implicit, realization of
the ground state of Matrix String Theory, that is compatible with the
dictionary (12) and (13).

Now one can write down the reduced planar Hamiltonian.
According to the fermionic-fluid picture of 
the ground state (25) the effective planar Matrix String Theory 
Hamiltonian is simply given by
$$
H_{r} = \int Dx_{i} Dp_{i}  D\phi^{T}  D\phi
( R\int d\alpha d\beta ({1 \over 2}p_{i}^{2}  +  
{1 \over 4}\{x_{i},x_{j}\}^{2} +  
 \phi^{T} \gamma_{i} \{\phi,x_{i}\})) 
\theta(e - H),     \eqno(26)
$$
or in terms of a fermionic functional $\Psi$ which describes the
fermionic nature of the vacuum  
$$
H_{r} = R\int d\alpha d\beta \int Dx_{i} D \phi^{T} 
D \phi 
({1 \over 2}{\delta {\Psi^{\dagger}} \over {\delta x_{i}}}
{\delta {\Psi} \over {\delta x_{i}}} +({1 \over 4}\{x_{i},x_{j}\}^{2}
+  \phi^{T} \gamma_{i} \{\phi,x_{i}\} - e)
\Psi^{\dagger} \Psi ).      \eqno(27)
$$
This formula can be understood as the usual expression for the
ground state energy, written in a
second quantized manner, after taking into account the fact that the 
ground
state of the planar theory is fermionic, as implied by (25).

Note that the above expressions for the effective Matrix String Theory
Hamiltonian contain functional integrals, the
fact which tells us that we are not dealing with an
ordinary field theory.

One could use the bosonic weight functional $\rho = \Psi^{\dagger} \Psi $ 
as a "collective functional"
in the spirit of [11]. Then the expression (26) could be interpreted
as the effective potential of the Das-Jevicki-Sakita collective-functional
 Hamiltonian. The minimum of the 
effective potential, which determines the ground state  of
Matrix String Theory, is in turn given by (25). 
Unfortunately, unlike in the one-matrix
model case [12], a simple explicit expression for the 
collective functional $\rho$ cannot be readily obtained.

\vskip .2in
\noindent{\bf 4. In Lieu Of Conclusion}
\vskip .2in

Matrix String Theory presents a plausible non-perturbative
formulation of string theory in which the number of degrees of 
freedom differs from ordinary quantum field theory, 
 yet because of the nature of the 
planar limit,  it also differs from ordinary quantum mechanics. Actually,
as this article attempts to show,
Matrix String Theory is not an ordinary quantum theory. That fact is 
indicated
by equations (5), (7) and (8) which define a Hamiltonian version of
"quansical" theory, or
planar quantum theory. (The ground state of the theory being given
by (25) and the effective planar Hamiltonian 
being given by (27).) This theory in turn is different from
Adler's generalized trace quantum dynamics, which can be also 
applied to the structure of Matrix String Theory [4].

The outlined "quansical"-theoretic structure is closely related 
 to Connes' non-commutative quantized calculus [13].
For example, consider the definition of 
Connes' non-commutative "quantum" derivative [13]
$$
d_{c}O = [F,O],
\eqno(28)
$$
where $F$ is a self-adjoint operator 
such that $F^{2}=1$ or $F={2 C -1}$, where
$C$ is a projection operator $C^{2} = C$. 
 (For example, $C=\psi^{\dagger} \psi$,
or $F=\psi^{\dagger} + \psi$,
where $\psi^{\dagger}$ and $\psi$ are fermionic as before.)
Hence, $d_{c}O$ anticommutes with $F$.
In principle, the "quansical" commutation relations (5) can be written as
$$
[[X_{i},P_{j}],X_{k}] = {1 \over 2} i \hbar \delta_{ij} d_{c}X_{k}.
\eqno(29)
$$
Then $P_{i}$ can be expressed in terms of $X_{j}$ and $d_{c}X_{k}$.
The resulting expression turns out to be unfortunately rather complicated.
(It is interesting to note, though, that in the
zero-dimensional case the nature of the operator 
$F$ is uniquely determined [13],
the operator $F$ being given by the familiar expression for the
Hilbert transform, which appears in the usual treatment of the one-matrix 
model [12], and the non-commutative 
"quantum" derivative being given by the following intuitively appealing
expression $d_{c}O f(s) = \int {{O(s) - O(t)} \over {s - t}} f(t) dt$, 
where
$f(s)$ is some suitable function.)
 
Perhaps even more important is the relation 
of (5) to non-commutative probability theory of
Voiculescu [14], especially the idea of Free Fock spaces, defined by the
action of "free" operators $a_{i}$, $a_{j}^{\dagger}$
 ($a_{i}a_{j}^{\dagger} = \delta_{ij}$, and
$a |0{\rangle} = 0$). As shown in [14],
the "quansical" commutation relations (5) can
be naturally obtained if appropriate operator representations of
$X_{i}$ and $P_{j}$ are given in terms of free operators
$a$ and $a^{\dagger}$. (In the present context this fact would imply that 
the
non-perturbative Matrix String 
Theory Fock space is a Free Fock space.) Note that 
this operator representation is quite different
from the one considered in section 3., which was an explicit fermionic
 representation
of the ground state compatible with the particular
 representation of the fundamental
non-commuting variables of Matrix String Theory.

A few words are in order about the nature of 
the above realization as compared to 
the old matrix model formulation of zero and one-dimensional 
non-perturbative string physics [15].
The two are indeed very similar in spirit. 
(The above formulation is in some sense an extension of [15].)
One obvious difference is the use of functionals (as opposed to functions
[15]) in the present approach.
Perhaps the most striking difference between the two approaches
 is the appearance of two extra 
continuous parameters ($\alpha$ and $\beta$) in the above 
formulation, 
compared to one extra parameter-dimension (the familiar
eigenvalues of the one-matrix model) of [15]. Unfortunately most of 
the above functional expressions
are still quite formal. The question of renormalization after appropriate
regularization has been completely ignored. One would expect the
appearance of one extra parameter (the renormalization 
scale $\lambda$) making the total number
of parameters in the above Hamiltonian formulation equal to three.
The true dynamics of Matrix String Theory would be then governed by a
Wilsonian non-perturbative RG equation describing a scale-by scale 
evolution of
the effective Matrix String Theory Hamiltonian (27). 

I close this article with a sketch of a possible 
speculative interpretation/visualization 
of the outlined realization of Matrix String Theory as a generalized
quantum theory.
The "quansical" commutation relations (5) appear natural from the
point of view of quantum cosmology: the projection operator 
$|0 {\rangle}{\langle}0|$
 nicely captures the classical features of a quantum system, such as the
Universe, that at some point during its evolution
becomes very large (that is - classical) 
as compared to some initial fundamental scale.
The fermionic ground state (the ground state of the Universe)
could be interpreted from the point of view of quantum information theory 
[16] 
and in accordance with Susskind's holographic principle [17]. 
A bit of information obtained through quantum measurements
is created/destroyed by the action of
creation/annihilation fermionic
operators that feature in the second quantized description of the
fermionic ground state. All information is stored at
the boundary of a region defining a fundamental "hole" (or "monad") of 
space
(a gauge invariant bound state of Matrix String Theory) inside which it is
in principle impossible to make any observations, due to the
non-commutative character of space within the "hole". 
The Fermi energy corresponds in this picture
to that fundamental energy scale above which it is in principle 
impossible to obtain information through quantum measurements.
Given such fermionic quantum-informational ground state of the Universe, 
the non-commutative coordinates that appear in (5) could be understood
as being dynamically induced through wave functional overlaps, or
in other words, as being the same infrared stable
 planar "gauge connections" that feature in the 
geometric phase of the fermionic ground state. 
\vskip.1in

I thank Shyamoli Chaudhuri, Joseph Polchinski and Branko Uro\v{s}evi\'{c} 
for 
useful comments on the initial manuscript. 
I am also grateful to Bunji Sakita for drawing my attention to his work 
with Kavalov, and I thank V. P. Nair for initial discussions.
I also wish to thank members of the University of Chicago Theory Group
 for their kind hospitality.
This work was fully and generously supported by 
the Joy K. Rosenthal Foundation.

\vskip.1in
{\bf References}
\item{1.} T. Banks, W. Fischler, S. Shenker, L. Susskind, Phys. Rev. D55 
(1997)
5112.
\item{2.} E. Witten, Nucl. Phys. BB460 (1995) 335.
\item{3.} For a review of D-brane physics see: J. Polchinski, S. 
Chaudhuri and
C. V. Johnson, {\it "Notes on D-branes"}, NSF-ITP-96-003, hep-th/9602052.
\item{4.} S. L. Adler, Nucl. Phys. B415 (1994) 195; hep-th/9703053.
\item{5.} G. 't Hooft, Nucl. Phys. B72 (1974) 461; E. Witten, 
Nucl. Phys. B160 (1979) 519; for a collections of papers on large N
methods consult {\it The large N Expansion in Quantum Field Theory 
and Statistical Mechanics}, eds. E. Brezin and S. Wadia, World Scientific,
1994. 
\item{6.} R. P. Feynman and A. R. Hibbs, 
{\it Quantum mechanics and Path Integrals}, McGraw-Hill, 1965.
\item{7.} J. Schwinger, {\it Quantum Kinematics and Dynamics},
 W. A. Benjamin, 1970.
\item{8.} D. Minic, {\it On The Planar Yang-Mills Theory}, May 1997.
\item{9.} K. Bardakci, Nucl. Phys. B178 (1980) 263; 
O. Haan, Z. Phys C6 (1980) 345; M. B. Halpern, Nucl. Phys. B188 (1981) 
61;  
M. B. Halpern and C. Schwarz, Phys. Rev. D24 (1981) 2146; 
A. Jevicki and N. Papanicolaou, Nucl. Phys. B171 (1980) 363;
A. Jevicki and H. Levine, Phys. Rev. Lett 44 (1980) 1443;
 Ann. Phys. 136 (1981) 113.
\item{10.} E. G. Floratos, J. Iliopoulos and G. Tiktopoulos, 
Phys. Lett. B217 (1989) 285; J. Hoppe, Ph. D. thesis (MIT, 1988); see also
A. Kavalov and B. Sakita, hep-th/9603024. 
\item{11.} B. Sakita, Phys. Rev. D21, (1980) 1067; 
A. Jevicki and B. Sakita, Nucl. Phys. B165 (1980) 511; 
Nucl. Phys. B185 (1981) 89; 
S. R. Das and A. Jevicki, Mod. Phys. Lett. A5 (1990) 1639.
\item{12.} E. Brezin, C. Itzykson, G. Parisi and J. B. Zuber, 
Comm. Math. Phys. 59 (1978) 35.
\item{13.} A. Connes, {\it Noncommutative Geometry}, Academic
Press, 1994.
\item{14.} D. V. Voiculescu, K. J. Dykema and A. Nica, {\it Free
Random Variables}, AMS, Providence 1992; 
M. Douglas, Phys. Lett. B344 (1995) 117; Nucl. Phys. Proc. Suppl. 41 
(1995) 66;
R. Gopakumar and D. Gross, Nuc. Phys. B451 (1995) 379.
\item{15.} D. Gross and A. Migdal, Phys. Rev. Lett. 64 (1990) 127;
M. Douglas and S. Shenker, Nucl. Phys. B335 (1990) 635; 
E. Brezin and V. Kazakov, Phys. Lett. B236 (1990) 144; for a highly
inspiring review consult: J. Polchinski {\it What is String Theory?}.
1994 Les Houches lectures, hep-th/9411028.
\item{16.} See J. A. Wheeler and W. Zurek, eds., {\it Quantum Theory and
Measurement}, Princeton University Press, 1983.
\item{17.} L. Susskind, J. Math. Phys. 36 (1995) 6377.

\end